# Epitaxial growth and antiferromagnetism of Sn-substituted perovskite iridate SrIr$_{0.8}$Sn$_{0.2}$O$_3$

Junyi Yang[1,*], Lin Hao[1,*], Qi Cui[2,3], Jiaqi Lin[4], Lukas Horak[5], Xuerong Liu[6], Lu Zhang[2], Huaixin Yang[2,3,7], Jenia Karapetrova[8], Jong-Woo Kim[8], Philip J. Ryan[8], Mark P. M. Dean[4], Jinguang Cheng[2,3,7], Jian Liu[1,*]

1. Department of Physics and Astronomy, University of Tennessee, Knoxville, Tennessee 37996, USA;
2. Beijing National Laboratory for Condensed Matter Physics and Institute of Physics, Chinese Academy of Sciences, Beijing 100190, China;
3. School of Physical Sciences, University of Chinese Academy of Sciences, Beijing 100190, China;
4. Department of Condensed Matter Physics and Materials Science, Brookhaven National Laboratory, Upton, New York 11973, USA;
5. Department of Condensed Matter Physics, Charles University, Ke Karlovu 5, 121 16 Prague, Czech Republic;
6. School of Physical Science and Technology, ShanghaiTech University, Shanghai 201210, China;
7. Songshan Lake Materials Laboratory, Dongguan, Guangdong 523808, China;
8. Advanced Photon Source, Argonne National Laboratory, Argonne, Illinois 60439, USA.

*To whom all correspondence should be addressed.

jyang43@vols.utk.edu; lhao3@utk.edu; jianliu@utk.edu

Abstract

5$d$ iridates have shown vast emergent phenomena due to a strong interplay among its lattice, charge and spin degrees of freedom, because of which the potential in spintronic application of the thin-film form is highly leveraged. Here we have epitaxially stabilized perovskite SrIr$_{0.8}$Sn$_{0.2}$O$_3$ on [001] SrTiO$_3$ substrates through pulsed laser deposition and systematically characterized the structural, electronic and magnetic properties. Physical properties measurements unravel an insulating ground state with a weak ferromagnetism in the compressively strained epitaxial film. The octahedral rotation pattern is identified by synchrotron x-ray diffraction, resolving a mix of $a^+b^-c^-$ and $a^-b^+c^-$ domains. X-ray magnetic resonant scattering directly demonstrates a G-type antiferromagnetic structure of the magnetic order and the spin canting nature of the weak ferromagnetism.

The combination of electron-electron interactions, spin-orbit coupling (SOC), and crystal field all at similar energy scales has raised increasing interests in discovering emergent phenomena in 5$d$ transition metal oxides, such as topological phases and unconventional magnetism [1-3]. One of the particular focuses is the so-called $J_{eff}$ = 1/2 electrons of iridates, highlighted by the Ruddlesden-Popper (RP) series $Sr_{n+1}Ir_nO_{3n+1}$ [3-13]. While the two-dimensional endmember $Sr_2IrO_4$ has an antiferromagnetic (AFM) insulating ground state [5,14], perovskite $SrIrO_3$ (SIO) at the three-dimensional limit shows a paramagnetic semi-metallic behavior [4,15-19]. The semi-metallicity is believed to be caused by the band-crossing at a Dirac nodal ring [18,20,21] protected by a nonsymmorphic crystalline symmetry [22-24]. With epitaxial constraint, it was demonstrated that the nonsymmorphic symmetry can be removed, but no sign of magnetic ordering has been found [19,23]. On the other hand, it was found that magnetic transitions could be triggered through partial chemical substitutions of the Ir site with various ions in perovskite SIO [25,26]. While the Ir valence is altered in many of these cases, one outstanding example is a near-room-temperature weak ferromagnetism that was discovered in isovalent nonmagnetic $Sn^{4+}$-substituted perovskite $SrIr_{1-x}Sn_xO_3$, which also becomes insulating in contrast to SIO [27,28]. While these emergent behaviors render this substitution series as promising functional magnetic materials, only polycrystalline samples can be synthesized in the bulk [27], which requires high pressure synthesis similar to the parent compound SIO [15,17,29]. The lack of single-crystalline samples hinders further exploration of the magnetic structure and the underlying mechanism of magnetism induced by the nonmagnetic substitutions. Furthermore, even though many SOC-driven phenomena reported in bulk iridates have significant implications for spintronics [30-33], achieving their thin film form is the first step towards device applications [34-38].

In this work, we show that high-quality epitaxial thin films of perovskite $SrIr_{0.8}Sn_{0.2}O_3$ (SISO) can be synthesized on $SrTiO_3$ (STO) single-crystal substrates. Resistivity and magnetic measurements show that the SISO films are insulators with a weak ferromagnetic (FM) transition, which are consistent with the bulk [27]. The spontaneous magnetization is found to be much larger within the film plane than along the out-of-plane direction. Meanwhile, the epitaxially stabilized structure allows us to perform a thorough investigation on the perovskite lattice distortion with synchrotron x-ray diffraction, from which two orthorhombic twin domains due to $IrO_6$ octahedral rotation were found. We further probed the underlying magnetic structure by performing x-ray magnetic resonant scattering experiment and directly identified a G-type AFM ground state, confirming the spin canting nature of the weak ferromagnetism.

SISO thin films were deposited from a polycrystalline target of nominal SISO composition [27] on STO (001)-oriented single crystal substrates by using a pulsed laser deposition system (KrF excimer laser) equipped with a reflection high-energy electron diffraction (RHEED) unit. During deposition, the substrate temperature and laser fluence were kept as 700 °C and 2 J/cm$^2$, respectively, with a constant oxygen pressure of 115 mTorr. Crystal structure and crystallinity of the thin films were investigated by x-ray diffraction (XRD) measurements on a Panalytical X'Pert MRD diffractometer using Cu Kα radiation (1.5406 Å). The surface quality was characterized by atomic force microscopy. The composition was checked through energy dispersive spectroscopy (EDS). Magnetic properties measurements were carried out on a Vibrating Sample Magnetometer (Quantum Design). Sample resistivity was measured by using the standard four-point probe on a Physical Property Measurement System (Quantum Design). Synchrotron x-ray diffraction measurements were carried out at room temperature at beamline 33-BM, and x-ray resonant magnetic scattering experiments were performed at beamline 6-ID-B at the Advanced Photon Source of the Argonne National Laboratory.

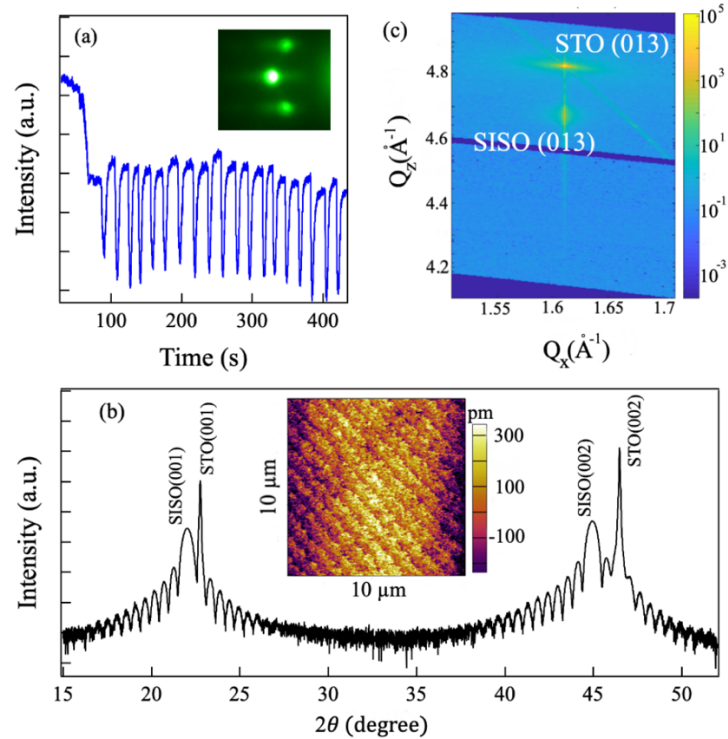

FIG. 1. (a) RHEED oscillation pattern during the thin film growth. The inset is the RHEED image after growth. (b) XRD $\theta - 2\theta$ scan along the STO (001) direction. The pseudo-cubic unit cell $a \times a \times c$ is used for the notation. The inset is the surface morphology map by atomic force microscopy. (c) Reciprocal space mapping around the SISO (013) plane.

The epitaxial deposition process is controlled by in-situ monitoring the RHEED intensity of the specular reflection. As shown in Fig. 1(a), we observed well-defined RHEED oscillations over the whole growth process, indicating a stable layer-by-layer growth mode of SISO on the STO substrate. The RHEED pattern recovers the bright spots of both specular and off-specular reflections lying on the zeroth *Laue* circle after the growth of each layer as demonstrated in the inset of Fig. 1(a), indicating a smooth surface. Shown in Fig. 1(b) is a representative XRD $\theta - 2\theta$ scan of the obtained SISO films along the STO (001) direction, from which clear film Bragg reflections are observed next to the STO substrate peaks. No impurity phase reflection is found, confirming the single perovskite phase of the samples. The out-of-plane lattice constant can be determined as 4.028 Å which is significantly larger than the bulk (~3.966 Å) [27]. The out-of-plane elongation can be explained as a direct response to the ~ –1.6% in-plane compressive strain of SISO when growing on STO (3.905 Å). In addition, high-contrast fringes are observed around the film reflections suggesting a coherent and sharp film-substrate interface. From the size of the fringes, we obtained a thickness of ~ 18 nm for this particular sample shown in Fig. 1(b), which is consistent with the estimation based on the number of RHEED oscillations, confirming the layer-by-layer growth control with a growth rate about one unit cell per 5 seconds. Clear terraces because of the substrate miscut can be seen on the sample surface, and the surface roughness with the terrace is ~ 0.1 nm (Fig. 1(b) inset). From reciprocal space mapping measurement, we confirmed that the SISO thin film has the same in-plane lattice parameter with that of the STO substrate and therefore is fully strained (Fig. 1(c)). The volume of the pseudocubic cell obtained from the measured lattice parameter is ~61.42 Å$^3$, which is ~1.6% smaller than the bulk [27] due to the compressive strain.

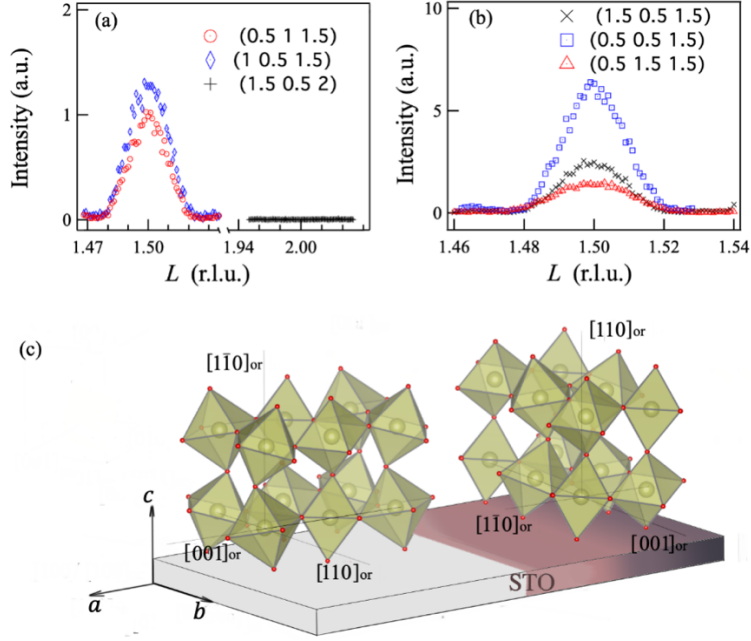

FIG. 2. Synchrotron XRD measurements. (a) In-phase rotation Bragg peaks. (1.5 0.5 2), (1 0.5 1.5) and (0.5 1 1.5) correspond to $c^+$, $a^+$ and $b^+$ octahedral rotations, respectively. (b) Out-of-phase rotation Bragg peaks. (0.5 0.5 1.5), (0.5 1.5 1.5) and (1.5 0.5 1.5) corresponds to $a^-/b^-$, $b^-/c^-$ and $a^-/c^-$ octahedral rotations, respectively. (c) Schematic diagrams of twin crystal domains of the SISO thin film. The grey zone denotes the $a^+b^-c^-$ rotation pattern while the brown zone represents the $a^-b^+c^-$ rotation pattern.

To further resolve the epitaxial crystalline structure of the SISO thin film beyond the lattice constants measurements, we investigated the octahedral rotation pattern via synchrotron XRD measurements. Bulk $SrIr_{1-x}Sn_xO_3$ was previously found to be orthorhombic with the *Pbnm* space group due to the $a^-a^-c^+$ perovskite octahedral rotation pattern [27], where the pseudocubic $c^+$-axis is parallel to the orthorhombic $[001]_{or}$ direction. The superscript "+ (–)" here from the Glazer notation [39] denotes an in-phase (out-of-phase) rotation manner of successive octahedra in a projection view. Since octahedral rotation shifts oxygen ions away from the centrosymmetric positions and expands the primitive unit cell, the rotation pattern can be characterized by a set of half-integer Bragg reflections. Following Glazer's method [39,40], we performed a broad survey to find the in-phase rotation axis of the SISO thin film (Fig. 2(a)). There is no observable peak at (1.5 0.5 2) which indicates that film $c$-axis is not an in-phase rotation axis. In contrast, a clear (1 0.5 1.5) peak is observed, suggesting that the *a*-axis is the in-phase rotation axis. Additionally, we also observed a (0.5 1 1.5) peak, pointing to a *b*-axis in-phase octahedral rotation. Since bulk orthorhombic cell has only one in-phase rotation axis, the coexistence of the two in-phase rotation axes suggests that the [001]$_{or}$ axis is within the film plane with a twin domain structure, as schematically shown in Fig. 2(c). In other words, the SISO thin film has a mixed $a^+b^-c^-$ and $a^-b^+c^-$ rotation pattern when grown on the cubic STO substrate. This scenario can be further verified by measuring the associated out-of-phase rotation peaks shown in Fig. 2(b). We found the (0.5 1.5 1.5) peak corresponding to $b^-/c^-$ rotation, the (1.5 0.5 1.5) peak corresponding to $a^-/c^-$ rotation, and the (0.5 0.5 1.5) peak corresponding to $a^-/b^-$ rotation. The observation of all these three peaks demonstrates out-of-phase rotations along all three axes, consistent with the picture of the 90°-rotated orthorhombic twin domains shown in Fig. 2(c). The rotation angles of the two out-of-phase axes cannot be quantified here but are likely to be different due to epitaxial strain [23,41].

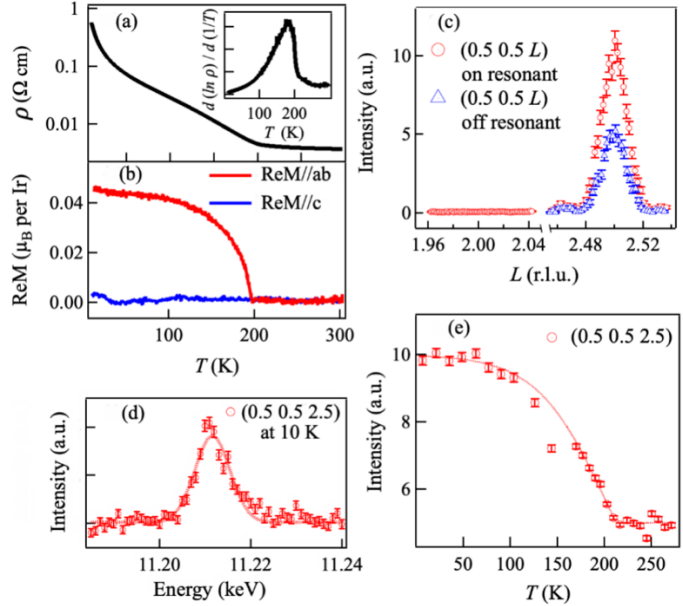

FIG. 3. (a) Representative temperature dependence of the SISO thin film resistivity. Inset shows the temperature dependence of $d(\ln \rho)/d(1/T)$. (b) Temperature dependence of remnant moment per Ir (ReM) within the *ab*-plane (red) and along the *c*-axis (blue). (c) Resonant x-ray magnetic scattering measurements (red circles) along the (0.5 0.5

$L$) truncation rod at 10 K (Ir $L_3$-edge). Data collected at off-resonant condition (blue triangles) is also shown for comparison. (d) Energy profile of the (0.5 0.5 2.5) Bragg reflection peak across the Ir $L_3$-edge at 10 K. The dashed line served as a guidance. (e) Temperature evolution of the magnetic Bragg peak (0.5 0.5 2.5) (Ir $L_3$-edge). The dashed line serves as a guideline. The error bars denote the statistic error.

Next, we show electronic and magnetic properties of the obtained SISO thin film. As shown in Fig. 3(a), the room-temperature resistivity of the SISO thin film is about 2 mΩ·cm, similar to the SIO thin films grown on STO [42,43]. However, in contrast to the weak temperature dependence of the semimetallic resistivity of SIO, the resistivity of SISO thin film increases by more than two orders of magnitude when cooling from room temperature to 10 K. This indicates an insulating nature of SISO, which is consistent with the bulk polycrystalline sample [27]. In addition, we also observed a clear resistivity kink around 200 K, which can be further seen in the plot of $d\ln(\rho)/d(1/T)$ as a function of temperature as a $\lambda$-shape cusp peaks at 185 K (inset of Fig. 3(a)). This behavior was also reported in the bulk and associated with a magnetic transition [27]. To shed light on the magnetic state of the SISO thin film, we measured the temperature dependent remnant magnetization. As shown in Fig. 3(b), the in-plane remnant magnetization (ReM//$ab$-plane) sharply increases when temperatures cooled down below 200 K, and saturates at ~ 0.043 $\mu_B$/Ir at the base temperature, close to the reported value in the bulk counterpart [27]. Meanwhile, no significant remnant magnetization along the out-of-plane direction (ReM//$c$-axis) was observed. The onset temperature of the weak FM transition is similar to the onset temperature of the resistivity kink, confirming a strong coupling between electronic transport and magnetic ordering in the SISO film.

The origin of the weak ferromagnetism of SISO is believed to be associated with spin canting of a G-type AFM order of the $J_{eff} = 1/2$ moments on Ir sites [27], but the AFM structure has yet to be resolved completely. While neutron scattering is often used to probe magnetic structure, the limited sample volume of such thin films with the small $J_{\text{eff}} = 1/2$ moment of the neutron-absorbing Ir ion precludes the feasibility of this route. Instead, we performed resonant x-ray magnetic scattering measurements around the Ir $L_3$-edge, which is an element-selective technique with resonant enhancement of the magnetic signal. The magnetic contribution to the scattering intensity was further extracted by suppressing the charge contribution through the ($\sigma - \pi$) channel of a pyrolytic graphite analyzer. Shown in Fig. 3(c) is a representative Bragg reflection with magnetic contribution found at (0.5 0.5 2.5) at 10 K. The peak intensity drops by about half when the photon energy is tuned away from the Ir $L_3$-edge, indicating significant resonance with the Ir $5d$ state. The off-resonance intensity at this position is a structural peak from the out-of-phase octahedral rotation discussed earlier and remains detected here because of a combination of its strong intensity and the incomplete efficiency of the analyzer in blocking the ($\sigma - \sigma$) channel. Since the octahedral rotation peak is not supposed to have structural contribution from the Ir sublattice, the observed resonance is likely to due to magnetic order. This is confirmed by the energy profile shown in Fig. 3(d), which maximizes the peak intensity at the energy slightly lower than the Ir $L_3$ white line with a typical magnetic line shape [5,14,44,45]. The fact that the magnetic Bragg peak appears at the (0.5 0.5 2.5) position directly demonstrates that Ir magnetic moment is antiferromagnetically coupled with all six nearest-neighboring sites, forming a G-type AFM structure. This can be further verified from the absence of any (0.5 0.5 *integer*) magnetic reflection as shown in Fig. 3(c). When warming the sample, the magnetic component starts to decrease above 100 K as shown in Fig. 3(e) and eventually disappears around 200 K, above which only the non-resonant structural component remains. This temperature dependence is consistent with that of the magnetization and resistivity kink, demonstrating a canted G-type AFM insulating ground state of SISO.

In the G-type AFM order, each Ir moment is antiparallel aligned with all six neighboring Ir moments regardless whether the neighbor's octahedron is rotated in-phase or out-of-phase. The antiferromagnetic interaction can be considered as Heisenberg-like superexchange, and it is not expected to be very sensitive to the $90^0$-rotated orthorhombic domain boundary, unlike the antiphase domains [46,47]. On the other hand, since the octahedral rotation introduces Dzyaloshinskii–Moriya interactions [48], magnetic anisotropy could be influenced and the AFM axis may change direction across the orthorhombic domain wall. In addition, given that the strong in-plane remnant magnetization is consistent with shape anisotropy of thin film, it is unclear whether the intrinsic magnetic anisotropy favors spin-canting in the plane or out of the plane. Finally, we noticed the AFM ordering temperature is lower than the bulk. We checked the Sn concentration of the film by EDS, which shows a 1:3 atomic ratio between Sn and Ir but with significant error due to the small film signal. Although the deviation from the nominal Sn concentration could be related to the nonequilibrium process of the deposition, this result suggests that the lower ordering temperature is not due to Sn-deficiency. Other effects, such as epitaxial strain, could be in play. All these open questions are interesting directions for future investigations, given the possibility of directly probing the AFM order in such thin films by resonant x-ray.

In conclusion, we have synthesized perovskite $SrIr_{0.8}Sn_{0.2}O_3$ epitaxial thin films on $SrTiO_3$ (001) substrate. Synchrotron x-ray diffraction unraveled two perpendicular in-phase octahedral rotation axes within the *ab*-plane, indicative of two structural domains of the orthorhombic cell. No in-phase rotation around the out-of-plane direction was observed. Physical properties measurements revealed an insulating state of the film, where the resistivity is enhanced upon a weak ferromagnetic transition, implying a close interplay between electronic and magnetic degrees of freedom. Stabilization of the epitaxial films enables investigation on the microscopic magnetic structure by utilizing resonant x-ray magnetic scattering, which demonstrates a G-type antiferromagnetic ground state underlying the weak ferromagnetism. While significant spin-canting has been well reported in a number of the AFM layered iridates due to the strong SOC and octahedral rotation [5,49-52], perovskite SISO expands this interesting material family as a 3D representative that can be epitaxially stabilized in the form of thin films. These results open the door to studying and engineering the emergent properties of magnetic perovskite iridate through epitaxy.

The authors acknowledge experimental assistance from H.D. Zhou, M. Koehler. J.L acknowledges support from the Science Alliance Joint Directed Research & Development Program and the Organized Research Unit Program at the University of Tennessee. This material is based upon work supported by the National Science Foundation under Grant No. DMR-1848269. L.H. acknowledges the support by the ERDF (project CZ.02.1.01/0.0/0.0/15_003/0000485) and the Grant Agency of the Czech Republic grant (14-37427 G). Work at Brookhaven National Laboratory was supported by the U.S. Department of Energy, Office of Science, Office of Basic Energy Sciences, under Contract No. DE-SC0012704 and Early Career No. 1047478. Use of the Advanced Photon Source, an Office of Science User Facility operated for the US DOE, OS by Argonne National Laboratory, was supported by the U. S. DOE under contract no. DE-AC02-06CH11357. J.G.C. is supported by the National Key R&D Program of China (Grants No. 2018YFA0305700), the National Natural Science Foundation of China (Grants No. 11874400), the Key Research Program of Frontier Sciences of the Chinese Academy of Sciences (Grants No. QYZDB-SSW-SLH013). X.R.L, H.X.Y and J.G.C. are also supported by the CAS